\documentclass[aps,preprint,showkeys,showpacs]{revtex4}
\usepackage{graphicx}
\begin{document}
\newcommand{\be}{\begin{equation}}
\newcommand{\ee}{\end{equation}}
\newcommand{\bea}{\begin{eqnarray}}
\newcommand{\eea}{\end{eqnarray}}

\title{Yang-Lee zeros of the one-dimensional $Q$-state Potts model}
\author{Seung-Yeon Kim\footnote{Electronic address: sykim@kias.re.kr}}
\affiliation{School of Computational Sciences,
Korea Institute for Advanced Study,\\
207-43 Cheongryangri-dong, Dongdaemun-gu, Seoul 130-722, Korea}

%\date{\today}

\begin{abstract}
The distributions of the Yang-Lee zeros of the ferromagnetic
and antiferromagnetic $Q$-state Potts models in one dimension
are studied for arbitrary $Q$ and temperature.
The Yang-Lee zeros of the Potts antiferromagnet have been
fully investigated for the first time.
The distributions of the Yang-Lee zeros
show a variety of different shapes.
Some of the Yang-Lee zeros lie on the positive real axis even for $T>0$.
For the ferromagnetic model this happens only for $Q<1$,
while there exist some zeros of the antiferromagnetic model
on the positive real axis both for $Q<1$ and for $Q>1$.
\end{abstract}

\pacs{05.50.+q; 64.60.Cn; 75.10.Hk}
\keywords{Potts model; Partition function zeros;
Ferromagnetic Yang-Lee zeros; Antiferromagnetic Yang-Lee zeros}

\maketitle

%%%%%%%%%%%%%%%%%%%%%%%%%%%%%%%%%%%%%%%%%%%%%%%%%%%%%%%%%%%%%%%%%%%%%%%%%%%

\section{introduction}

The $Q$-state Potts model \cite{potts,wu82,wu84,baxter,martin}
is a generalization of the Ising ($Q=2$) model.
The $Q$-state Potts model exhibits
a rich variety of critical behavior
and is very fertile ground for the analytical and numerical investigations
of first- and second-order phase transitions.
The Potts model is also related to other outstanding
problems in physics and mathematics.
Fortuin and Kasteleyn \cite{kasteleyn,fortuin} have shown that
the $Q$-state Potts model in the limit $Q\to1$ defines
the problem of bond percolation.
They \cite{fortuin} also showed that the problem of resistor network
is related to a $Q=0$ limit of the partition function of the Potts model.
In addition, the zero-state Potts model describes the statistics
of treelike percolation \cite{stephen}, and is equivalent to the undirected
Abelian sandpile model \cite{majumdar}.
The $Q={1\over2}$ state Potts model has a connection to a dilute spin glass
\cite{aharony}.
The $Q$-state Potts model with $0\le Q\le1$ describes
transitions in the gelation and vulcanization processes
of branched polymers \cite{lubensky}.
The partition function of the Potts model is also known as
the Tutte dichromatic polynomial \cite{tutte}
or the Whitney rank function \cite{whitney}
in graph theory and combinatorics of mathematics.

By introducing the concept of the zeros of the partition function
in the {\it complex} magnetic-field plane,
Yang and Lee \cite{yang} proposed a mechanism
for the occurrence of phase transitions in the thermodynamic limit
and yielded a new insight into the unsolved problem of the ferromagnetic
Ising model in an arbitrary nonzero external magnetic field.
It has been known
\cite{yang,lee,itzykson,creswick97,creswick98,creswick99,janke01,janke02a,janke02b,alves,kromhout}
that the distribution of the zeros of a model
determines its critical behavior.
Lee and Yang \cite{lee} also
formulated the celebrated circle theorem which states that
the partition function zeros of the Ising ferromagnet lie on the unit circle
in the complex magnetic-field ($x=e^{\beta H}$) plane.
However, for the ferromagnetic $Q$-state Potts model with $Q>2$
the Yang-Lee zeros in the complex $x$ plane
lie close to, but not on, the unit circle with the two
exceptions of the critical point $x=1$ ($H=0$) itself and the
zeros in the limit $T=0$ \cite{creswick99,kim98,kim99,kim00,kim02}.
It has been shown \cite{kim00,kim02}
that the distributions of the ferromagnetic Yang-Lee zeros for $Q>1$
have similar properties independent of dimension.

Recently, the exact results on the Yang-Lee zeros of the ferromagnetic
Potts model have been found using the one-dimensional model.
Mittag and Stephen \cite{mittag} studied the Yang-Lee zeros
of the three-state Potts ferromagnet in one dimension.
Glumac and Uzelac \cite{glumac} found the eigenvalues of the transfer matrix
of the one-dimensional Potts model for general $Q$.
In particular, they have shown that
the ferromagnetic Yang-Lee zeros can lie on the real axis for $Q<1$.
However, the properties of the Yang-Lee zeros of
the {\it antiferromagnetic} Potts model have never been known
except for the one-dimensional Ising antiferromagnet \cite{yang2}.
In this paper, we study the ferromagnetic and antiferromagnetic
Yang-Lee zeros of the one-dimensional $Q$-state Potts model.

%%%%%%%%%%%%%%%%%%%%%%%%%%%%%%%%%%%%%%%%%%%%%%%%%%%%%%%%%%%%%%%%%%%%%%%%%%%%%%%%%

\section{partition function}

The $Q$-state Potts model in an external magnetic field $H_q$
on a lattice $G$ with $N_s$ sites and $N_b$ bonds
is defined by the Hamiltonian
\be
{\cal H}_Q=-J\sum_{\langle i,j\rangle}\delta(\sigma_i,\sigma_j)
-H_q\sum_k\delta(\sigma_k,q),
\ee
where $J$ is the coupling constant (ferromagnetic model for $J>0$
and antiferromagnetic model for $J<0$),
$\langle i,j\rangle$ indicates a sum over nearest-neighbor pairs,
$\sigma_i=1,2,...,Q$,
and $q$ is a fixed integer between 1 and $Q$.
The partition function of the model is
\be
Z_Q=\sum_{\{ \sigma_n \}} e^{-\beta{\cal H}_Q},
\ee
where $\{\sigma_n\}$ denotes a sum over $Q^{N_s}$ possible spin
configurations and $\beta=(k_B T)^{-1}$.
The partition function can be written as
\be
Z(a,x,Q)=\sum_{E=0}^{N_b}\sum_{M=0}^{N_s}\Omega_Q(E,M) a^E x^M,
\ee
where $a=y^{-1}=e^{\beta J}$, $x=e^{\beta H_q}$,
$E$ and $M$ are positive integers $0\le E\le N_b$ and $0\le M\le N_s$,
respectively, and $\Omega_Q(E,M)$ is the number of states
with fixed $E$ and fixed $M$.
The states with $E=0$ ($E=N_b$) correspond to the antiferromagnetic
(ferromagnetic) ground states.
For ferromagnetic interaction $J>0$ the physical interval is
$1\le a\le\infty$ ($\infty\ge T\ge0$), whereas
for antiferromagnetic interaction $J<0$ the physical interval
$0\le a\le1$ ($0\le T\le\infty$).
The parameter $Q$ enters the Potts model as an integer.
However, the study of the $Q$-state Potts model has been extended to
continuous $Q$ due to the Fortuin-Kasteleyn representation
of the partition function \cite{kasteleyn,fortuin}
and its extension \cite{blote}.

For the one-dimensional Potts model in an external field the
eigenvalues of the transfer matrix were found by Glumac and
Uzelac \cite{glumac}. The eigenvalues are $\lambda_\pm=(A\pm i B)/2$,
where $A=a(1+x)+Q-2$ and $B=-i\sqrt{[a(1-x)+Q-2]^2+4(Q-1)x}$,
and $\lambda_0=a-1$ which is $(Q-2)$-fold degenerate.
The partition function of the one-dimensional model ($N=N_s=N_b$) is given by
\be
Z_N(a,x,Q)=\lambda_+^N+\lambda_-^N+(Q-2)\lambda_0^N.
\ee

%%%%%%%%%%%%%%%%%%%%%%%%%%%%%%%%%%%%%%%%%%%%%%%%%%%%%%%%%%%%%%%%%%%%%%%%%

\section{partition function zeros}

When $\lambda_+$ and $\lambda_-$ are two dominant eigenvalues,
the partition function becomes
\be
Z_N\simeq\lambda_+^N+\lambda_-^N
\ee
for large $N$.
If we define $A=2C\cos\psi$ and $B=2C\sin\psi$,
where $C=\sqrt{(a-1)(a+Q-1)x}$, then $\lambda_\pm=C\exp(\pm i\psi)$,
and the partition function is
\be
Z_N=2C^N \cos N\psi.
\ee
The zeros of the partition function are then given by
\be
\psi=\psi_k={2k+1\over2N}\pi,\ \ \ \ \ k=0,1,2,...,N-1.
\ee
In the thermodynamic limit the locus of the partition function zeros is determined
by the solution of
\be
A=2C\cos\psi,
\ee
where $0\le\psi\le\pi$.
In the special case $Q=2$ the contribution by the eigenvalue $\lambda_0$
disappears from the partition function, Eq.~(4),
and the equation (8) determines all the locus even for finite systems.
From Eq.~(8) the locus of the Yang-Lee zeros for any $Q$ is obtained
to be
\be
x_1(\psi)={1\over a^2}\bigg[\sqrt{f_1}\cos\psi + i\sqrt{f_2}\bigg]^2,
\ee
where $f_1=(a-1)(a+Q-1)$ and
$f_2=f_1\sin^2\psi+Q-1$. The edge zeros of $x_1(\psi)$ are given
by
\be
x_\pm={1\over a^2}\bigg[\sqrt{(a-1)(a+Q-1)}\pm\sqrt{1-Q}\bigg]^2
\ee
from $x_1(0)$ and $x_1(\pi)$. If $f_1>0$ and $f_2>0$ or $f_1<0$ and
$f_2<0$, it is easily verified that
\be
|x_1(\psi)|=\sqrt{x_+x_-}=\bigg|{a+Q-2\over a}\bigg|.
\ee
From Eq.~(11) we see that the
zeros of $x_1(\psi)$ lie on a circle in the complex $x$ plane. The
one point of the circle $x_1(\psi)$,
\be
x_1\bigg({\pi\over2}\bigg)=-{a+Q-2\over a},
\ee
always lies on the real axis.
However, if $f_1>0$ and $f_2<0$ or $f_1<0$ and $f_2>0$,
the zeros of $x_1(\psi)$ lie on the real axis.
Recently, Ghulghazaryan et al. \cite{ghulghazaryan} tried to understand
the properties of Eq.~(9) in the antiferromganetic ($a\le1$) region.

On the other hand, when $\lambda_+$ and $\lambda_0$ are two
dominant eigenvalues, the partition function can be written as
\be
Z_N\simeq\lambda_+^N+(Q-2)\lambda_0^N
\ee
for large $N$. The
partition function zeros are then determined by
\be
{\lambda_+\over\lambda_0}=(2-Q)^{1/N}\exp(i\phi),
\ee
where
\be
\phi=\phi_k={2\pi k\over N},\ \ \ \ \ k=0,1,2,...,N-1.
\ee
In the thermodynamic limit the locus of the partition function zeros is
determined by the solution of
\be a^2x^2+(Q-1)x-a x A+(a-1)A e^{i\phi}-(a-1)^2e^{2i\phi}=0,
\ee
where $0\le\phi\le2\pi$.
The equation (16) also determines the locus of the zeros when
$\lambda_-$ and $\lambda_0$ are two dominant eigenvalues.
Eq.~(16) gives the second locus of the Yang-Lee
zeros
\be
x_2(\phi)={e^{i\phi}[a+Q-2-(a-1)e^{i\phi}]\over a+Q-1-a e^{i\phi}},
\ee
which is the unit circle in the limit $a\to\infty$
and a type of lima\c{c}on of Pascal for $a=0$. The two points of
the locus $x_2(\phi)$,
\be
x_2(0)=1
\ee
and
\be
x_2(\pi)=-{2a+Q-3\over2a+Q-1},
\ee
can be lie on the real axis. From
$|\lambda_\pm|=|\lambda_0|$ we also obtain
\be
x_\ast={a-1\over a+Q-1}.
\ee
The second locus, Eq.~(17), of the Yang-Lee zeros, although it is very
important in understanding the $Q$-state Potts model, has never been
considered in the literature until now.

%%%%%%%%%%%%%%%%%%%%%%%%%%%%%%%%%%%%%%%%%%%%%%%%%%%%%%%%%%%%%%%%%%%%%%%%%%%%

\section{ferromagnetic ($\lowercase{a}\ge1$) Yang-Lee zeros}

In this section we study the Yang-Lee zeros of the ferromagnetic
$Q$-state Potts model for $a\ge1$.

\subsection{$Q>1$}

In this case $f_1$ and $f_2$ are always positive and the Yang-Lee zeros
$x_1(\psi)$ lie on a circle where the eigenvalues satisfy
\be
{|\lambda_\pm|\over|\lambda_0|}=\sqrt{(a+Q-1)(a+Q-2)\over a(a-1)}\ \ >1,
\ee
which implies that the locus $x_2(\phi)$ does not appear.
For $1<Q<2$ the zeros lie inside the unit circle while
for $Q>2$ the zeros lie outside the unit circle, as shown in \cite{kim00,glumac}.
In the special case $Q=2$ we of course find that $|x_1(\psi)|=1$,
as proved by Lee and Yang \cite{lee}.

The argument $\theta$ of $x_1(\psi)$ $(=|x_1(\psi)|e^{i\theta(\psi)}$) is given by
\be
\cos{\theta\over2}=\sqrt{(a-1)(a+Q-1)\over a(a+Q-2)}\cos\psi.
\ee
As $a\to\infty$ ($T\to0$), the zeros approach the unit circle.
For $a=1$ ($T=\infty$), $\cos(\theta/2)=0$,
so $\theta(\psi)=\pi$ and all the zeros lie at $1-Q$.
As $Q\to\infty$, the radius $|x_1(\psi)|$ increases without bound.
The Yang-Lee edge zero for $T>0$ is given by
\be
\theta_0=2\cos^{-1}\sqrt{(a-1)(a+Q-1)\over a(a+Q-2)}>0,
\ee
while $\theta_0=0$ at $T=0$.
Therefore, we conclude that no zero lie on the positive real axis for any $T>0$.

%%%%%%%%%%%%%%%%%%%%%%%%%%%%%%%%%%%%%%%%%%%%%%%%%%%%%%%%%%%%%%%%%%%%%%%%%%%%%%%%%

\subsection{$Q<1$}

At $a=1$ ($T=\infty$), all the Yang-Lee zeros lie at the point $1-Q$
on the positive real axis.
For $1<a\le1-{Q\over2}+{\sqrt{Q}\over2}$, all the zeros
lie on the positive real axis between $x_-$ and $x_+$ ($0<x_-<x_+<1$),
as pointed out by Glumac and Uzelac \cite{glumac,ghulghazaryan,monroe}.
For $a>1-{Q\over2}+{\sqrt{Q}\over2}$, the locus consists of
the loop $x_2(\phi)$ ($\phi_\ast\le\phi\le2\pi-\phi_\ast$) and
the line $x_1(\psi)$ ($0\le\psi\le\psi_\ast$) on the positive real axis
between $x_\ast=x_1(\psi_\ast)$ and $x_+=x_1(0)$ ($0<x_\ast<x_+<1$).
The loop $x_2$ meets with the line $x_1$ at the point
\be
x_\ast=x_2(\phi_\ast)=x_2(2\pi-\phi_\ast),
\ee
where $|\lambda_+|=|\lambda_-|=|\lambda_0|$.
The loop $x_2(\phi)$ cuts the real axis at two points
$x_2(\pi)$ and $x_\ast$ ($>x_2(\pi)$).
The sign of $x_2(\pi)$ is positive for $1-{Q\over2}+{\sqrt{Q}\over2}<a<{3-Q\over2}$
and negative for $a>{3-Q\over2}$.
Figure 1 shows the locus of the Yang-Lee zeros for $Q={1\over2}$ and
$a={6\over5}$ from which we obtain $x_2(\pi)={1\over19}$,
$x_\ast={2\over7}$, and $x_+=0.8119$.

At $Q=0$, the line $x_1(\psi)$ shrink to the point
\be
x_\ast=x_+=x_2(0)=1,
\ee
and all the zeros lie on the loop $x_2(\phi)$ ($0\le\phi\le2\pi$)
for $a>1$.
As $Q\to1$, two points $x_\ast$ and $x_+$ approach $(a-1)/a$ together,
and the line $x_1(\psi)$ disappears.
At $a=\infty$ ($T=0$), the line $x_1(\psi)$ again shrink to the point
$x_\ast=x_+=1$, and the loop $x_2(\phi)$ becomes
the unit circle centered at the origin.

%%%%%%%%%%%%%%%%%%%%%%%%%%%%%%%%%%%%%%%%%%%%%%%%%%%%%%%%%%%%%%%%%%%%%%%%%%%%

\section{antiferromagnetic ($0\le\lowercase{a}\le1$) Yang-Lee zeros}

In this section we investigate the Yang-Lee zeros of the antiferromagnetic
$Q$-state Potts model for $0\le a\le1$.

\subsection{$Q>1$}

Because $f_1$ is always negative, the zeros of $x_1(\psi)$ lie on a circle
if $f_2<0$ ($\psi_0<\psi<\pi-\psi_0$) and on the real axis
if $f_2>0$ ($0\le\psi<\psi_0$ or $\pi-\psi_0<\psi\le\pi$).
For $a<1-{Q\over2}$, the locus consists of the line $x_1(\psi)$
($0\le\psi\le\psi_0$ and $\pi-\psi_0\le\psi\le\psi_\ast$) on the negative
real axis between $x_+=x_1(0)$ and $x_\ast=x_1(\psi_\ast)$ ($x_+<x_\ast<0$),
the circle $x_1(\psi)$ ($\psi_0\le\psi\le\pi-\psi_0$)
with the radius $|x_1(\psi)|$, and the loop $x_2(\phi)$
($0\le\phi\le\phi_\ast$ and $2\pi-\phi_\ast\le\phi\le2\pi$),
inside the circle $x_1$. The loop $x_2$ again meets with the line $x_1$ at
the point
\be
x_\ast=x_2(\phi_\ast)=x_2(2\pi-\phi_\ast),
\ee
where $|\lambda_+|=|\lambda_-|=|\lambda_0|$.
The circle cuts the real axis at two points $x_1^0$ and
$x_1\big({\pi\over2}\big)$ ($=-x_1^0>1$),
where the point $x_1^0$ is defined by
\be
x_1^0=x_1(\psi_0)=x_1(\pi-\psi_0)={a+Q-2\over a}.
\ee
Similarly, the loop also cuts the real axis at two points
$x_\ast$ ($<0$) and $x_2(0)$ ($=1$).
At $a=0$ ($T=0$), the circle $x_1(\psi)$ disappears for $1<Q<2$, and
the locus consists of the line $x_1(\psi)$ on the
real axis between $-\infty$ and $x_\ast$ ($<0$) and the loop $x_2(\phi)$.
For example, figure 2 shows the locus for $Q={11\over10}$ and $a={1\over10}$.
In this case we obtain $x_+=-54.83$, $x_1^0=-8$, $x_\ast=-{9\over2}$,
the radius $|x_1|=8$ for the circle, $\psi_0=48.19^\circ$,
$\psi_\ast=134.07^\circ$, and $\phi_\ast=46.02^\circ$.

At $a=1-{Q\over2}$, two points $x_\ast$ and $x_1^0$ on the real axis meet at
\be
x_\ast=x_1^0=-1,
\ee
other two points $x_1({\pi\over2})$ and $x_2(0)$ on the real axis also meet at
\be
x_1\Big({\pi\over2}\Big)=x_2(0)=1,
\ee
and the circle $x_1(\psi)$ and the loop $x_2(\phi)$
become the identical locus as the unit circle.
On this unit circle, three eigenvalues have the same magnitude
\be
|\lambda_+|=|\lambda_-|=|\lambda_0|={Q\over2}.
\ee
Therefore, the locus consists of the line $x_1(\psi)$ on the
real axis between $x_+$ and $x_\ast$ and the circle.

In the region $1-{Q\over2}<a<1-{Q\over2}+{\sqrt{Q}\over2}$,
the circle $x_1(\psi)$ disappears, and
the line $x_1(\psi)$ ($0\le\psi\le\psi_\ast$) on the negative real axis
between $x_+$ and $x_\ast$ ($x_+<x_\ast<0$) again meets with
the loop $x_2(\phi)$ ($\phi_\ast\le\phi\le2\pi-\phi_\ast$) at the point $x_\ast$.
The loop cuts the real axis at two points $x_\ast$ and $x_2(\pi)$.
The sign of $x_2(\pi)$ is positive ($0<x_2(\pi)<1$)
for $1-{Q\over2}<a<{3-Q\over2}$ and negative for
${3-Q\over2}<a<1-{Q\over2}+{\sqrt{Q}\over2}$.
Figure 3 shows the locus of the Yang-Lee zeros for $Q=3$ and
$a={1\over10}$ which give $x_2(\pi)=-{1\over11}$, $x_\ast=-{3\over7}$,
and $x_+=-777.84$.
At $a=0$ ($T=0$), the locus still consists of the line $x_1(\psi)$ on the
real axis between $-\infty$ and $x_\ast$ ($<0$)
and the loop $x_2(\phi)$ for $2<Q<4$.
At $T=0$, the sign of $x_2(\pi)$ is positive ($0\le x_2(\pi)<1$)
for $2<Q\le3$ and negative for $3<Q<4$.
As $Q\to1$, two points $x_+$ and $x_\ast$ approach $(a-1)/a$ together,
and the line $x_1(\psi)$ disappears.

At $a=1-{Q\over2}+{\sqrt{Q}\over2}$, $x_\ast=x_-$, and the loop $x_2(\phi)$ shrinks
to the point $x_\ast$. For $a>1-{Q\over2}+{\sqrt{Q}\over2}$,
the loop disappears, and the only locus is the line $x_1(\psi)$
on the negative real axis
between $x_+$ and $x_-$ ($x_+<x_-<0$).
As $Q\to\infty$, $x_\pm\to-\infty$ for $a<1$.
At $a=1$ ($T=\infty$), two edge zeros $x_+$ and $x_-$ meet, and
the line $x_1(\psi)$ shrinks to the point
\be
x_+=x_-=1-Q
\ee
for $Q>1$.
In the special case $Q=2$, $f_1<0$ and $f_2>0$, and
the line $x_1(\psi)$ is the only locus of the Yang-Lee zeros.
Therefore, all the Yang-Lee zeros
of the antiferromagnetic Ising model lie on the negative real axis
between $x_+$ and $x_-$ ($>x_+$) for $a<1$, as shown by Yang \cite{yang2}.
At $a=0$ ($T=0$), all the zeros lie on the negative real axis
between two edge zeros,
\be
x_a=-\infty
\ee
and
\be
x_b={(Q-2)^2\over4(1-Q)} \ \ \ \ \ (x_b\le0),
\ee
for $Q=2$ and $Q\ge4$.

%%%%%%%%%%%%%%%%%%%%%%%%%%%%%%%%%%%%%%%%%%%%%%%%%%%%%%%%%%%%%%%%%%%%%%%%%%%%%%%%

\subsection{$Q<1$}

For $0\le a<1-{Q\over2}-{\sqrt{Q}\over2}$, the locus consists of
the loop $x_2(\phi)$ ($0\le\phi\le\phi_\ast$ and $2\pi-\phi_\ast\le\phi\le2\pi$)
and the line $x_1(\psi)$ ($0\le\psi\le\psi_\ast$) on the positive real axis
between $x_\ast=x_1(\psi_\ast)$ and $x_+=x_1(0)$ ($1<x_\ast<x_+$).
The loop $x_2$ cuts the real axis
at two points $x_2(0)$ ($=1$) and $x_\ast$ ($>1$).
The loop again meets with the line $x_1$ at the point
\be
x_\ast=x_2(\phi_\ast)=x_2(2\pi-\phi_\ast),
\ee
where $|\lambda_+|=|\lambda_-|=|\lambda_0|$.
At $Q=0$, two points $x_2(0)$ and $x_\ast$ meet, the loop $x_2(\phi)$
shrinks to the point $x_2(0)=x_\ast$, and all the Yang-Lee zeros lie
on the positive real axis between $x_-$ ($=x_\ast=1$) and $x_+$ ($\ge x_-$)
for $a\le1$.

At $a=1-{Q\over2}-{\sqrt{Q}\over2}$, $x_\ast=x_+$, and the line $x_1(\psi)$
shrinks to the point $x_\ast$.
In the region $1-{Q\over2}-{\sqrt{Q}\over2}<a<1-{Q\over2}$,
the line disappears, and the locations of Yang-Lee zeros are completely
determined by the locus $x_2(\phi)$ ($0\le\phi\le2\pi$)
which cuts the real axis at two points $x_2(0)$ ($=1$) and $x_2(\pi)$ ($>1$).
The shape of the locus $x_2(\phi)$ is changed from a waterdrop-like shape for
$1-{Q\over2}-{\sqrt{Q}\over2}<a<1-Q$ through
a circle with center $1\over1-Q$ and radius $Q\over1-Q$ for $a=1-Q$ to
a crescent-like shape for $1-Q<a<1-{Q\over2}$.
Figure 4 shows the locus of the Yang-Lee zeros for $Q={9\over10}$ and
$a={2\over5}$ which give $x_2(\pi)={13\over7}$.

At $a=1-{Q\over2}$, the loci $x_1(\psi)$ and $x_2(\phi)$
become identical to be the unit circle on which
three eigenvalues again have the same magnitude
\be
|\lambda_+|=|\lambda_-|=|\lambda_0|={Q\over2}.
\ee
For $a>1-{Q\over2}$, all the Yang-Lee zeros lie on the circle $x_1(\psi)$
($0\le\psi\le\pi$) which cuts the positive real axis at the point
$x_1({\pi\over2})$ ($0<x_1({\pi\over2})<1$).
At $a=1$ ($T=\infty$), this circle shrinks to the point $x_1(\psi)=1-Q$
for $Q<1$.

%%%%%%%%%%%%%%%%%%%%%%%%%%%%%%%%%%%%%%%%%%%%%%%%%%%%%%%%%%%%%%%%%%%%%%%%%%%%%%

\section{one-state (Q=1) Potts model}

At $Q=1$, the partition function becomes
\be
Z_N(a,x,Q=1)=(ax)^N.
\ee
Therefore, all the Yang-Lee zeros lie on the point $x=0$.
On the other hand, in the limit $Q\to1$,
the line $x_1(\psi)$ disappears, and
the circle $x_1(\psi)$ is the uniform distribution of the zeros
because the circle approaches
\be
x_1(\psi)=\bigg|{a-1\over a}\bigg|e^{2i\psi}.
\ee
The loop $x_2(\phi)$ also approaches
\be
x_2(\phi)={a-1\over a}e^{i\phi},
\ee
which is the same as the circle $x_1(\psi)$.

%%%%%%%%%%%%%%%%%%%%%%%%%%%%%%%%%%%%%%%%%%%%%%%%%%%%%%%%%%%%%%%%%%%%%%%%%%

\section{conclusion}

We have studied the interesting properties of the ferromagnetic
and antiferromagnetic Yang-Lee zeros of the one-dimensional $Q$-state
Potts model for arbitrary $Q$ and temperature.
It has been shown that the distributions of the Yang-Lee zeros
have a variety of different shapes.
In particular, the antiferromagnetic
Yang-Lee zeros of the Potts model have been fully investigated
for the first time.
One of the most interesting results is
that some of the Yang-Lee zeros
lie on the positive real axis even for $T>0$.
For the ferromagnetic Yang-Lee zeros this happens only for $Q<1$,
while there exist some Yang-Lee zeros of the antiferromagnetic Potts model
on the positive real axis both for $Q<1$ and for $Q>1$.
The results obtained from the one-dimensional Potts model may be
considered as a road to the full understanding
of the Yang-Lee zeros in higher dimensions in that for $Q>1$
the distributions of the ferromagnetic Yang-Lee zeros
have similar properties independent of dimension \cite{kim00,kim02}.

%%%%%%%%%%%%%%%%%%%%%%%%%%%%%%%%%%%%%%%%%%%%%%%%%%%%%%%%%%%%%%%%%%%%%%%%%%

\begin{acknowledgments}
The author is grateful to Prof. R.J. Creswick for very useful discussions.
\end{acknowledgments}

%%%%%%%%%%%%%%%%%%%%%%%%%%%%%%%%%%%%%%%%%%%%%%%%%%%%%%%%%%%%%%%%%%%%%%%%%%

%%%%%%%%%%%%%%%%%%%%%%%%%%%%%%%%%%%%%%%%%%%%%%%%%%%%%%%%%%%%%%%%%%%%%%%%%%
\newpage

\begin{figure}
\includegraphics[width=16cm]{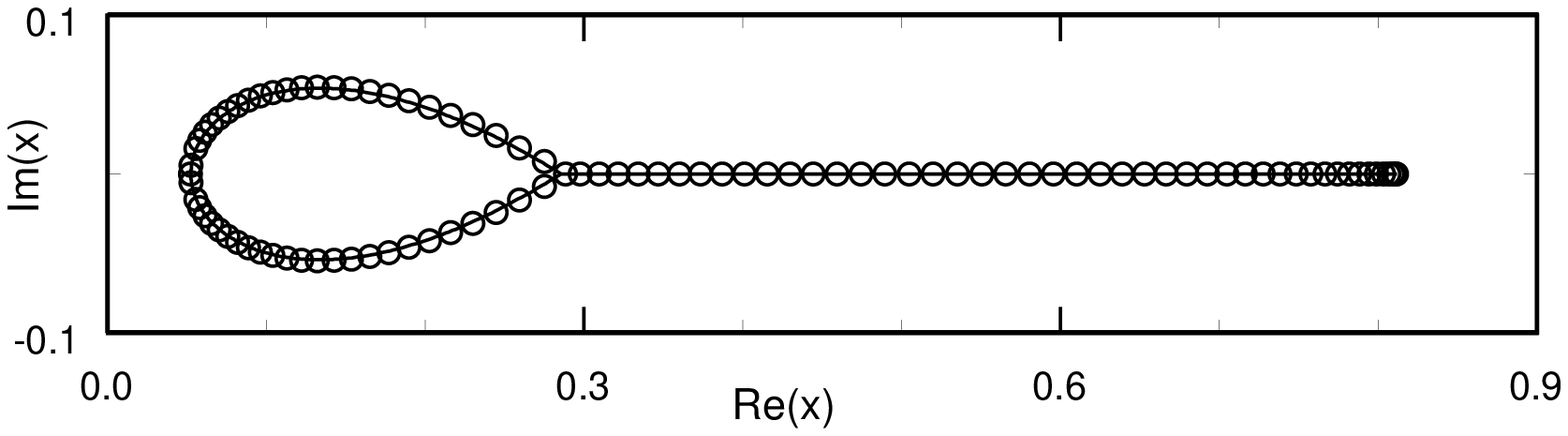}
\caption{The locus of the Yang-Lee zeros in the complex
$x=e^{\beta H_q}$ plane for $Q={1\over2}$ and $a=e^{\beta J}={6\over5}$.
For comparison, the zeros for a finite-size system ($N=100$)
are also shown (open circles).}
\end{figure}

\begin{figure}
\includegraphics[width=16cm]{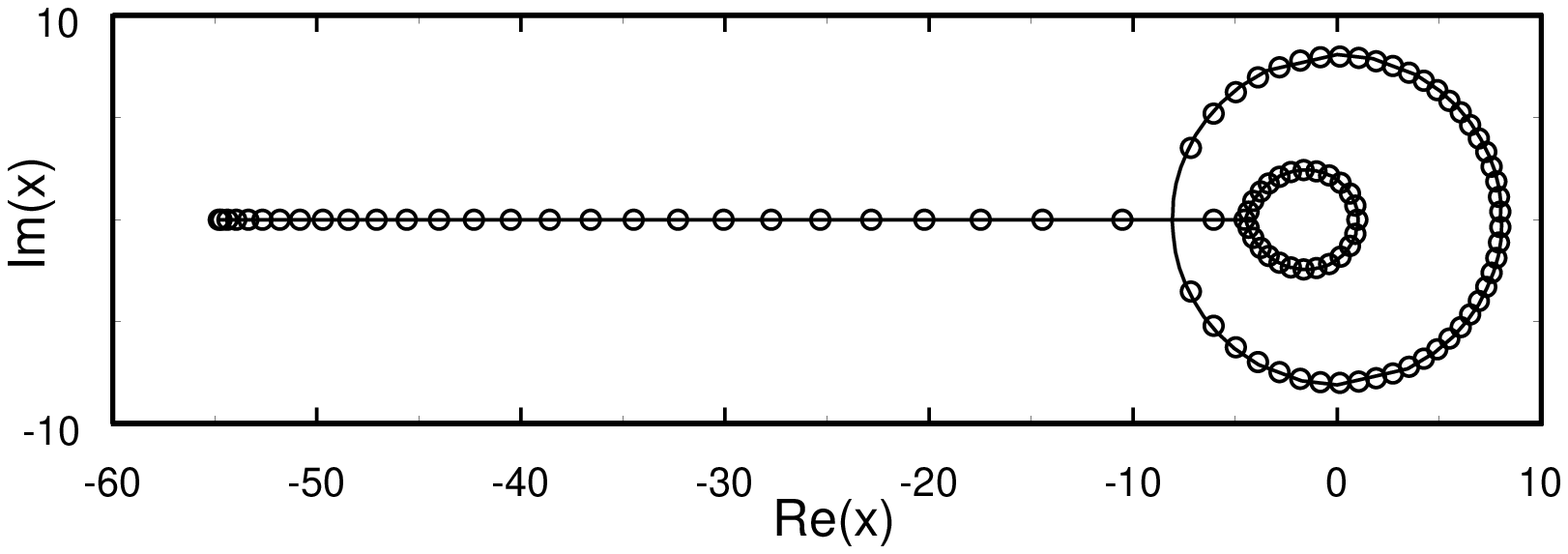}
\caption{The locus of the Yang-Lee zeros for $Q={11\over10}$ and $a={1\over10}$.}
\end{figure}

\begin{figure}
\includegraphics[width=16cm]{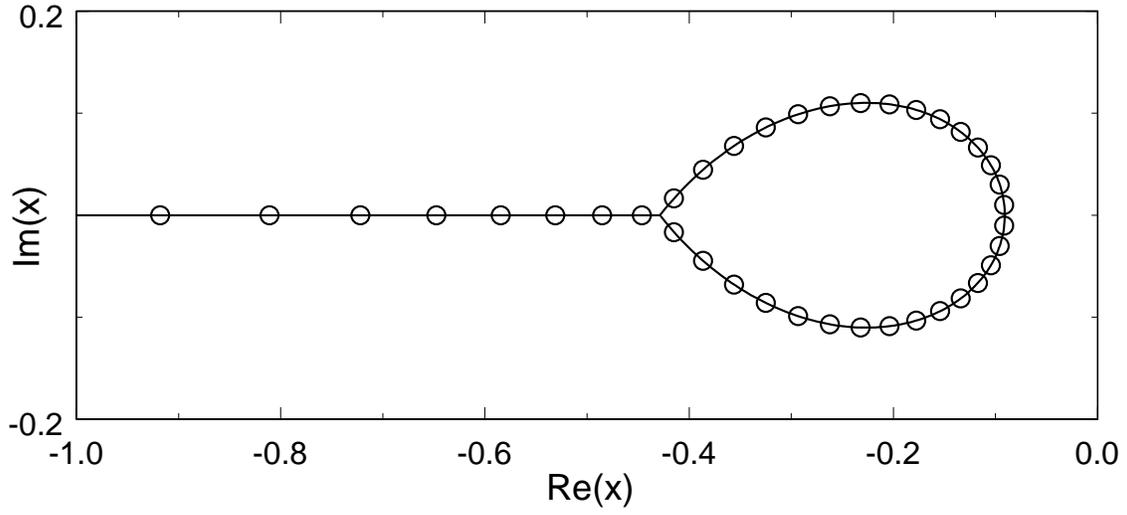}
\caption{The locus of the Yang-Lee zeros for $Q=3$ and $a={1\over10}$.
The zeros on the real axis lie between $x_\ast=-{3\over7}$ and $x_+=-777.84$.
Most of them are omitted in the figure.}
\end{figure}

\begin{figure}
\includegraphics[width=10cm]{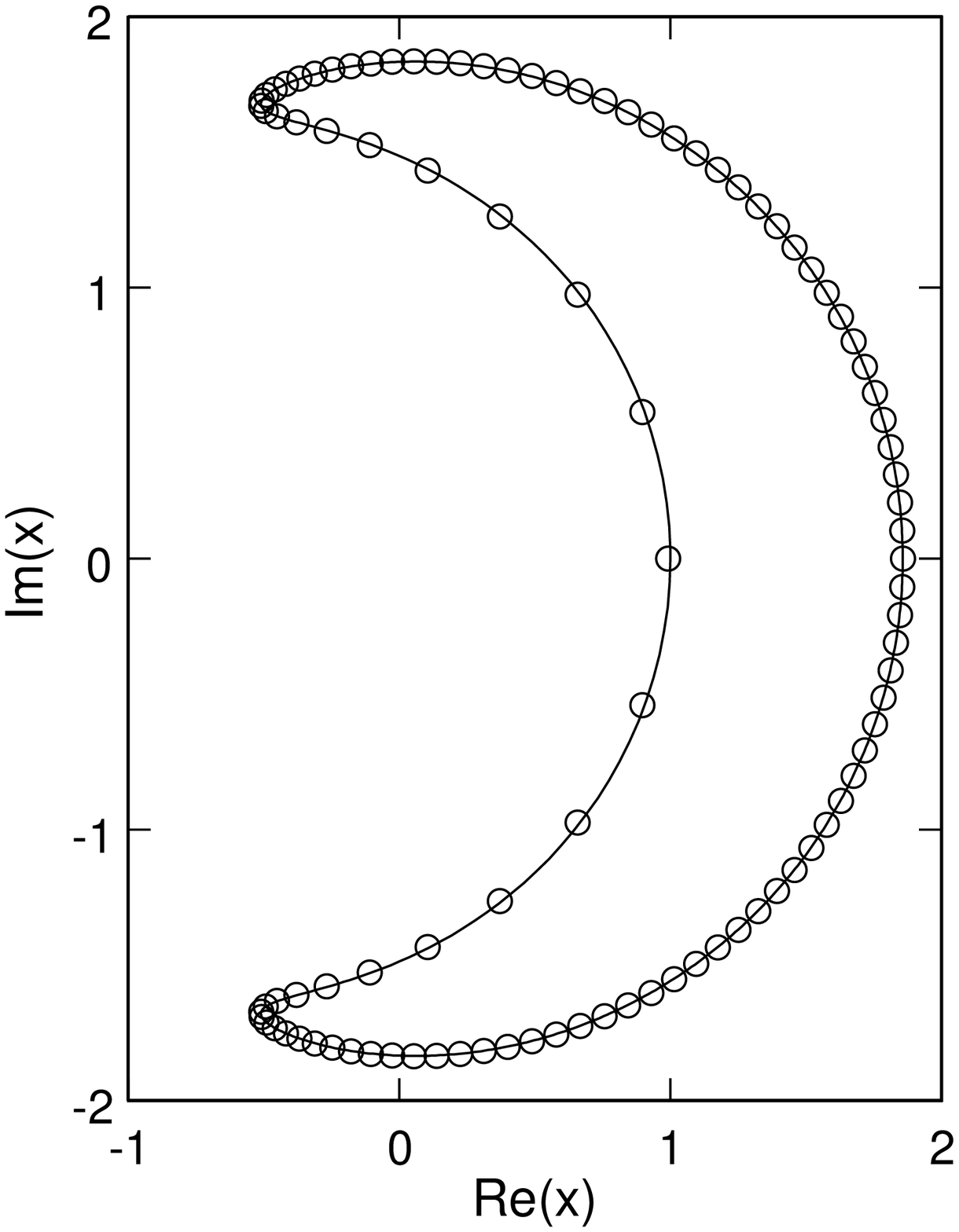}
\caption{The locus of the Yang-Lee zeros for $Q={9\over10}$ and $a={2\over5}$.}
\end{figure}

%%%%%%%%%%%%%%%%%%%%%%%%%%%%%%%%%%%%%%%%%%%%%%%%%%%%%%%%%%%%%%%%%%%%%%%%%%%%%%%

\end{document}